\documentclass[12pt]{article}
\usepackage{graphicx}
\usepackage[small,bf]{caption}

\catcode`\@=11 \long\def\@makefntext#1{ \protect\noindent \hbox to
3.2pt {\hskip-.9pt
$^{{\ninerm\@thefnmark}}$\hfil}#1\hfill}                

\def\@makefnmark{\hbox to 0pt{$^{\@thefnmark}$\hss}}  

\def\ps@myheadings{\let\@mkboth\@gobbletwo
\def\@oddhead{\hbox{}
\rightmark\hfil\ninerm\thepage}
\def\@oddfoot{}\def\@evenhead{\ninerm\thepage\hfil
\leftmark\hbox{}}\def\@evenfoot{}
\def\sectionmark##1{}\def\subsectionmark##1{}}

\renewcommand{\thefootnote}{\fnsymbol{footnote}}

\newcounter{sectionc}\newcounter{subsectionc}\newcounter{subsubsectionc}
\renewcommand{\section}[1] {\vspace*{0.6cm}\addtocounter{sectionc}{1}
\setcounter{subsectionc}{0}\setcounter{subsubsectionc}{0}\noindent
        {\normalsize\bf\thesectionc. #1}\par\vspace*{0.4cm}}
\renewcommand{\subsection}[1] {\vspace*{0.6cm}\addtocounter{subsectionc}{1}
        \setcounter{subsubsectionc}{0}\noindent
        {\normalsize\it\thesectionc.\thesubsectionc. #1}\par\vspace*{0.4cm}}
\renewcommand{\subsubsection}[1]
{\vspace*{0.6cm}\addtocounter{subsubsectionc}{1}
        \noindent
{\normalsize\rm\thesectionc.\thesubsectionc.\thesubsubsectionc.
        #1}\par\vspace*{0.4cm}}

\renewenvironment{thebibliography}[1]
        {\begin{list}{\arabic{enumi}.}
        {\usecounter{enumi}\setlength{\parsep}{0pt}
\setlength{\leftmargin 0.52cm}{\rightmargin 0pt}
         \setlength{\itemsep}{0pt} \settowidth
        {\labelwidth}{#1.}\sloppy}}{\end{list}}

\newcounter{itemlistc}
\newcounter{romanlistc}
\newcounter{alphlistc}
\newcounter{arabiclistc}

\def\@citex[#1]#2{\if@filesw\immediate\write\@auxout
        {\string\citation{#2}}\fi
\def\@citea{}\@cite{\@for\@citeb:=#2\do
        {\@citea\def\@citea{,}\@ifundefined
        {b@\@citeb}{{\bf ?}\@warning
        {Citation `\@citeb' on page \thepage \space undefined}}
        {\csname b@\@citeb\endcsname}}}{#1}}

\newif\if@cghi
\def\cite{\@cghitrue\@ifnextchar [{\@tempswatrue
        \@citex}{\@tempswafalse\@citex[]}}
\def\citelow{\@cghifalse\@ifnextchar [{\@tempswatrue
        \@citex}{\@tempswafalse\@citex[]}}
\def\@cite#1#2{{$\null^{#1}$\if@tempswa\typeout
        {IJCGA warning: optional citation argument
        ignored: `#2'} \fi}}

 1 
scaled\magstep 1  1
 
scaled\magstephalf

 \font\ninerm=cmr9

\newcommand{\beq}{\begin{equation}}
\newcommand{\eeq}{\end{equation}}

\newcommand\mathC{\mkern1mu\raise2.2pt\hbox{$\scriptscriptstyle|$}
                {\mkern-7mu\rm C}}

\begin{document}

\hfill \vspace*{1cm}

\centerline{\normalsize\bf MODIFIED NEWTONIAN DYNAMICS AS A
PREDICTION} \centerline{\normalsize\bf OF GENERAL RELATIVITY}

\vspace*{0.6cm} \centerline{\footnotesize SABBIR~A.~RAHMAN}
\vspace*{0.2cm} 
\vspace*{0.1cm} \centerline{\footnotesize E-mail:
sarahman@alum.mit.edu} \vspace*{0.9cm}

{\centering{\begin{minipage}{12.2truecm}\footnotesize\baselineskip=12pt
\noindent\centerline{\footnotesize ABSTRACT} \vspace*{0.1cm}
\parindent=0pt

We consider a simple model of the physical vacuum as a self-gravitating relativistic
fluid. Proceeding in a step-by-step manner, we are able to show that the equations of
classical electrodynamics follow if the electromagnetic four-potential is associated
with the four-momentum of a space-filling fluid of neutral spinors which we identify with
neutrinos and antineutrinos. Charged particles, which we identify with electrons
and positrons, act as sinks for the fluid and have the structure of the maximal fast
Kerr solution. Electromagnetic waves are described by oscillations in the fluid and
interactions between charges occur via the exchange of photons, which have the
structure of entwined neutrino-antineutrino pairs that form twisted closed loops
in spacetime connecting the charges. The model predicts that antimatter has negative
mass, and that neutrinos are matter-antimatter dipoles. Together these suffice
to explain the presence of modified Newtonian dynamics as a
gravitational polarisation effect.

\end{minipage}\par}}
\vspace*{0.6cm}

\normalsize\baselineskip=15pt \setcounter{footnote}{0}
\renewcommand{\thefootnote}{\alph{footnote}}    

\section{Introduction}

In a recent paper\cite{Blanchet06}, Blanchet showed that if there
were to exist a space-filling `aether' consisting of dark matter
particles which take the form of matter-antimatter dipoles, then
this would satisfactorily explain the existence of modified
Newtonian dynamics (MOND) as a simple gravitational polarisation
effect, in complete analogy with the polarisation of dielectrics in
classical electrodynamics.

In this paper, we begin by treating the physical vacuum as a featureless relativistic
continuum, and by proceeding in a step-by-step manner, are able not only
to derive classical electrodynamics essentially from first principles, but are
also able to show that the model gives rise to precisely the scenario
described by Blanchet, and therefore that MOND is a non-trivial
prediction of general relativity. In particular, our model predicts that
antimatter has negative mass and that neutrinos take the
form of matter-antimatter dipoles.

The structure of our paper is as follows. We begin in \S 2 with two
basic premises, namely (i) that the theory of relativity holds, and
(ii) that the physical vacuum is a featureless relativistic
continuum in motion. In \S 3 we show that it is possible to describe
the whole of classical electrodynamics in terms of the motion of a
two-component relativistic fluid where each component is a
time-reversed version of the other. In \S 4 we show that the
presence of this two component fluid is a prediction of general
relativity if charged particles have the structure of fast Kerr
black holes, and that the fluid particles are chargeless neutral
spinors which can be identified with neutrinos. We then show that
these neutrinos have precisely the properties required to explain
the occurrence of modified Newtonian dynamics as a gravitational
polarisation effect, and hence that MOND is a consequence of Einstein's
general theory of relativity. In \S 5 we briefly recall some speculations
that have been made about how the presence of antigravity might help to
resolve a number of outstanding issues in cosmology. We end in \S 6 with a summary and
discussion of our results.

We assume a metric with signature $(+, -, -, -)$, and follow the
conventions of Jackson\cite{JDJ98} throughout.

\section{Spacetime and the Physical Vacuum}

In this section we establish the reference system and the
coordinates that will be used to describe the dynamics of the
vacuum, and we show how Maxwell-like equations appear as identities
simply as a consequence of assuming that the underlying spacetime is
Lorentzian rather than Galilean.

\subsection{Coordinates and Reference Frames}

We begin our investigation with the assumption that the physical
vacuum is nothing but a featureless, space-filling, continuous
relativistic fluid (i.e. a relativistic continuum), whose properties
are described completely by its motion throughout (Minkowski)
spacetime. In particular, we  make no prior assumptions about
possible substructure or the mass density of the vacuum, so that the
only physical dimensions initially entering our discussion are those of length
and time.

Let us consider an {\it arbitrary} relativistic inertial frame of
reference in $M^{3,1}$ with 4-coordinates $x^\mu = (ct, x, y, z)$,
so that the spacetime partial derivatives are given by $\partial^\mu
= ({1\over c} {\partial\over\partial t}, - {\bf\nabla})$ and
$\partial_\mu = ({1\over c} {\partial\over\partial t}, {\bf\nabla})$
respectively. It is important to note that the forthcoming analysis
will be completely independent of the particular frame of reference
used.

Let $\tau$ be the proper time in this inertial frame, and let ${\bf
r}$ denote the 3-position $(x, y, z)$ of a point in the continuum.
Considering the instantaneous motion at proper time $\tau$ of the
continuum at any point ${\bf r}$, the 3-velocity of either component
of the continuum at that point as measured by the inertial frame is,
\begin{equation}
{\bf v} = {d{\bf r}\over dt}\,,\label{eqn:velocity}
\end{equation}
\noindent where $t$ is the time as measured by a clock moving with
the continuum. We can therefore define the interval,
\begin{equation}
ds^2 \equiv c^2 d\tau^2 = c^2 dt^2 - dx^2 - dy^2 -
dz^2\,.\label{eqn:interval}
\end{equation}
\noindent Similarly, we can define a 4-velocity vector field
describing the motion of the continuum as,
\begin{equation}
u^\mu = {dx^\mu\over d\tau} = \left(c {dt\over d\tau}, {d{\bf
r}\over d\tau}\right) = (c \gamma,
\gamma\bf{v})\,,\label{eqn:fourvel}
\end{equation}
\noindent where $\gamma = (1 - v^2/c^2)^{-1/2}$ is the Lorentz
factor at each point, and ${\bf v}$ is considered now as a 3-velocity
vector field. This 4-velocity clearly satisfies,
\begin{equation}
u^\mu u_\mu = c^2, \qquad u^{\mu, \nu} u_\mu = 0\,,\label{eqn:sol}
\end{equation}
\noindent where partial derivatives ${\partial^\nu u^\mu}$ are
written as $u^{\mu, \nu}$ for convenience.

\subsection{Maxwell-Like Equations}

\noindent Let us define the tensor $f^{\mu\nu}$ as the
antisymmetrised derivative of the 4-velocity,
\begin{equation}
f^{\mu\nu} = u^{\nu,\mu} - u^{\mu,\nu}\,.\label{eqn:twotensor}
\end{equation}
Then $f^{\mu\nu}$ satisfies the Jacobi identity,
\begin{equation}
f^{\mu\nu,\lambda} + f^{\nu\lambda,\mu} + f^{\lambda\mu,\nu} =
0\,,\label{eqn:jactwotensor}
\end{equation}
and we can define a 4-vector $j^\mu$ proportional to the divergence
of $f^{\mu\nu}$,
\begin{equation}
j^\mu = {c\over4\pi}
\partial_\nu f^{\mu\nu}\,,\label{eqn:fourtens}
\end{equation}
which has vanishing 4-divergence on account of the antisymmetry of
$f^{\mu\nu}$,
\begin{equation}
\partial_\mu j^\mu = 0\,.\label{eqn:vandiv}
\end{equation}

Equations (\ref{eqn:jactwotensor}) and (\ref{eqn:fourtens}) are
reminiscent of the homogeneous and inhomogeneous Maxwell's
equations, respectively.

\section{Classical Electrodynamics as Relativistic Fluid Dynamics}

Given the appearance of Maxwell-like equations
(\ref{eqn:jactwotensor}) and (\ref{eqn:fourtens}) it is natural to
ask whether our simple model of the physical vacuum can account for
classical electrodynamics. We show here that this is indeed the case
if the continuum fluid consists of two components which are
matter-antimatter conjugates of each other.

\subsection{The Continuum Gauge}

The first step would be to associate the electromagnetic 4-potential
$A^\mu$ with the 4-velocity of the continuum,

\begin{equation}
A^\mu=ku^\mu = (\phi, {\bf A}) = (kc\gamma, k\gamma{\bf
v})\,.\label{eqn:fourpottry}
\end{equation}
where $k$ is a positive dimensionful constant included to ensure
consistency of units on both sides. However the scalar potential
$\phi=kc\gamma$ would then be restricted to positive values,
resulting in an asymmetry between the descriptions of positive and
negative charges\footnote{For example, one can have
$\phi=kc\gamma=q/r$ for the scalar potential of a positive charge
but not $\phi=kc\gamma=-q/r$ for a negative charge.}.

For a charge-symmetric description of electrodynamics it is
necessary to split the electromagnetic potential 4-vector $A^\mu$
into the average of two components $A_+^\mu$ and $A_-^\mu$ which we
identify with two independent continuum 4-velocities $u_+^\mu$ and
$u_-^\mu$,

\begin{equation}
A^\mu = {1\over2}(A_+^\mu + A_-^\mu)\,,\quad\hbox{where}\quad A_+^\mu = ku_+^\mu = (\phi_+, {\bf A}_+)\,,\quad A_-^\mu = -ku_-^\mu = (\phi_-,
{\bf A}_-)\,.\label{eqn:fourpot}
\end{equation}

A charge-symmetric description of electrodynamics therefore requires
that the vacuum be a continuum consisting of two components\footnote{Although the introduction of two continuum
components may seem slightly {\it ad hoc} at present, the opposite
time signatures of their contributions to the 4-potential means that
$u_+^\mu$ and $u_-^\mu$ are associated with the motion of fluid
particles and antiparticles respectively, whose existence will be
shown in \S 4 to be a necessary consequence of general relativity.
The opposite sign of their respective contributions is due to the
opposite direction of propagation in time of the particles and
antiparticles. The fluid particles themselves can be identified with
neutrinos, which will be seen to be responsible both for cold dark
matter and for the emergence of modified Newtonian dynamics.} each in
motion which are related by reversal of time
signature, so that the 4-velocity in (\ref{eqn:fourpottry}) may be written\footnote{The negative contribution of the four-velocity $u_-^\mu$ may seem unphysical here, but will be found to be associated with the negative mass of antimatter when the 4-potential $A^\mu$ is eventually interpreted as the 4-momentum of the fluid.},

\begin{equation}
u^\mu={1\over2}(u_+^\mu-u_-^\mu)\,.\label{eqn:fourvelexp}
\end{equation}

\noindent The condition (\ref{eqn:sol}) implies the following covariant constraint
for both $A_+^\mu$ and $A_-^\mu$,

\begin{equation}
A_+^\mu A_{+ \mu} = A_-^\mu A_{- \mu} = k^2c^2\,.\label{eqn:gauge}
\end{equation}

We will refer to conditions (\ref{eqn:fourpot}) and
(\ref{eqn:gauge}) as the `continuum gauge'. This is a non-standard
choice of gauge, and we will demonstrate its consistency in \S 3.2
where we show that any electromagnetic field configuration can be
described uniquely by a potential 4-vector field with the form of
(\ref{eqn:fourpot}) satisfying the continuum gauge conditions.

The antisymmetric field-strength tensor can now be defined as,

\begin{equation}
F^{\mu\nu} = A^{\nu, \mu} - A^{\mu, \nu}\ \sim ({\bf E}, {\bf
B})\,.\label{eqn:stresstensor}
\end{equation}

Other standard properties now follow in the usual way. From the
definition (\ref{eqn:stresstensor}), $F^{\mu\nu}$ satisfies the
Jacobi identity,

\begin{equation}
F^{\mu\nu, \lambda} + F^{\nu\lambda, \mu} + F^{\lambda\mu, \nu} =
0\,,\label{eqn:jacobi}
\end{equation}

\noindent and this is just the covariant form of the homogeneous
Maxwell's equations. One can \emph{define} the 4-current as the
4-divergence of the field-strength tensor,

\begin{equation}
J^\mu = {c\over4\pi} F^{\mu\nu}_{,\nu}\ = (c \rho, {\bf
j})\,,\label{eqn:current}
\end{equation}

\noindent and this is the covariant form of the inhomogeneous
Maxwell's equations. Charge conservation is guaranteed by the
antisymmetry of the field-strength tensor. The covariant Lorentz
force equation takes the following form,

\begin{equation}
{dV^\mu\over d\tau} = {Q\over M c} F^{\mu\nu}
V_\nu\,,\label{eqn:lorentz}
\end{equation}

\noindent where $Q$, $M$ and $V^\mu = (c \gamma_V, \gamma_V {\bf
V})$ are the charge, mass and 4-velocity vector of the observed
particles. This cannot be derived directly from the definition of
the 4-potential, and must be considered for now as an auxiliary
constraint.

The charge 4-velocity $V^\mu$ and scalar charge $Q$ are related to
the 4-current density $J^\mu$ through the following equation,

\begin{equation}
J^\mu = Q V^\mu\,,\qquad\hbox{where}\quad V^\mu V_\mu = c^2\,,\quad
V^0\geq c\,.\label{eqn:charge}
\end{equation}

The constraint on $V^\mu$ allows us to separate the 4-current
uniquely into the charge and its 4-velocity. Indeed we have,

\begin{equation}
Q = \hbox{sgn}(J^0) \cdot \left({1\over c^2} J^\mu
J_\mu\right)^{1/2}\,,\label{eqn:zerocomp}
\end{equation}

\noindent where the sign of the 0-component of the 4-current appears
to ensure that the 0-component $V^0$ of the charge 4-velocity is
positive. The gauge based upon a single 4-vector field was precisely
that introduced by Dirac in his classical model of the
electron\cite{Dirac51a}, and it is noteworthy that he was also led
to speculate that this 4-velocity field described the motion of a
real, physical, `aether'\cite{Dirac51b}. The form of the charge
4-velocity in terms of the continuum 4-velocity now follows directly
from (\ref{eqn:charge}).

Besides the mass $M$ which is determined by initial conditions, each
of the terms in (\ref{eqn:lorentz}) may be written in terms of the
$4$-velocities $u_+^\mu$ and $u_-^\mu$. From the definitions of
$F^{\mu\nu}$, $J^\mu$, $Q$ and $V^\mu$, we find that the Lorentz
force equation (\ref{eqn:lorentz}) translates into a complicated
third order partial differential equation constraining the
$4$-velocities. The conservation of mass follows from the continuity
equation for mass density,

\begin{equation}
(M V^\mu)_{,\mu} = 0\,,\label{eqn:masscons}
\end{equation}

\noindent which is ensured if the flow of mass density follows the
flow of charge density. We will see later that the Lorentz force
equation follows from the fluid dynamical interactions between
sources and/or sinks, and this will complete our picture of classical
electrodynamics in this gauge.

\subsection{The Consistency of the Continuum Gauge}

We have identified the components $A_+^\mu$ and $A_-^\mu$ of the
4-potential with the 4-velocities $u_+^\mu$ and $u_-^\mu$ of the
continuum satisfying the conditions (\ref{eqn:fourpot}) and
(\ref{eqn:gauge}), and have referred to this gauge choice as the
`continuum gauge'. It is not obvious that this gauge choice can be
applied consistently to all electromagnetic field configurations, so
we demonstrate its consistency here, and give explicit solutions for
the point charge and the plane electromagnetic wave.

In order to prove consistency, it is necessary to find a
decomposition of the 4-potential as the difference of two 4-velocity
fields satisfying equations (\ref{eqn:fourpot}) and
(\ref{eqn:gauge}) simultaneously. Using the notation of
(\ref{eqn:fourvel}), we therefore need to find, given any
4-potential $A^\mu=(\phi,{\bf A})$ defined up to a gauge
transformation $A^\mu\rightarrow A^\mu+\partial^\mu\psi$, two
3-velocity fields ${\bf v}_+$ and ${\bf v}_-$ satisfying the
following conditions,

\begin{equation}
{\phi\over kc} = \gamma_+-\gamma_-\,,\qquad{{\bf A}\over k} =
\gamma_+{\bf v}_+-\gamma_-{\bf v}_-\,.\label{eqn:gammadiff}
\end{equation}

The second of these equations is a simple geometrical vector
identity, and it is clear that any solution set for $(\gamma_+{\bf
v}_+, \gamma_-{\bf v}_-)$ will form a surface of revolution about
the axis defined by ${\bf A}$. To find the solution surface
explicitly for a given $(\phi,{\bf A})$, it is convenient to take
the origin to lie at ${\bf A}/2k$, and to use polar coordinates
$(r,\theta)$ in any plane containing ${\bf A}$, where
$r\in[0,\infty]$ is the radial distance from the origin and
$\theta\in[0,\pi]$ is the angle made with respect to the direction
of ${\bf A}$. Note the following simple chain of identities,

\begin{equation}
\gamma = {1\over\sqrt{1-{v^2\over c^2}}} \Rightarrow v =
c\,\sqrt{1-{1\over\gamma^2}} \Rightarrow \gamma v =
c\sqrt{\gamma^2-1} \Rightarrow \gamma = \sqrt{1+\left({\gamma v\over
c}\right)^2}\,,\label{eqn:gammidents}
\end{equation}
so that from (\ref{eqn:gammadiff}) we have,
\begin{equation}
{\phi\over kc} = \sqrt{1+\left({\gamma_+v_+\over c}\right)^2} -
\sqrt{1+\left({\gamma_-v_-\over
c}\right)^2}\,.\label{eqn:gammadiff2}
\end{equation}
Applying standard trigonometric identities to our geometrical
picture, we obtain,
\begin{equation}
(\gamma_+v_+)^2 = r^2+A^2/4k^2+{Ar\over k}
\cos\theta\,,\qquad(\gamma_-v_-)^2 = r^2+A^2/4k^2-{Ar\over k}
\cos\theta\,,\label{eqn:gv}
\end{equation}
so that the set of solutions on the plane in question is determined
by the condition,
\begin{equation}
\phi = \sqrt{A^2/4+Akr\cos\theta+k^2(r^2+c^2)} -
\sqrt{A^2/4-Akr\cos\theta+k^2(r^2+c^2)}\,.\label{eqn:phisol}
\end{equation}
Note that given any solution for $(\phi,{\bf A})$, a solution for
$(-\phi,{\bf A})$ is obtained by letting
$\theta\rightarrow\pi-\theta$. Note also (i) that $\phi=0$ whenever
$\theta=\pi/2$ including when $r=0$, (ii) that for a given value of
$r$ the magnitude of $\phi$ is maximum when $\theta=0$, (iii) that
for $\theta=0$, $\phi$ is a monotonically increasing function of
$r$, and (iv) that $\phi\rightarrow A\cos\theta$ as
$r\rightarrow\infty$.

In conclusion, for a given value of $A=|{\bf A}|$, equations
(\ref{eqn:gammadiff}) will have solutions whenever $|\phi| \leq A$.
In the special case $\phi=0$ the solution surface for $\gamma_+{\bf
v}_+$ is just the plane perpendicular to ${\bf A}$ passing through
the point ${\bf A}/2k$, throughout which $|{\bf v}_+|=|{\bf v}_-|$,
and $|\gamma_+{\bf v}_+|\geq A/2k$. For other values of $|\phi| \leq
A$ the solutions form a paraboloid-like surface of revolution about
the ${\bf A}$ axis. The sign of $\phi$ determines which side of the
$\theta=\pm\pi/2$ plane the solution surface lies.

It is always possible to choose the function $\psi$ defining the
choice of gauge in such a way that $\phi=0$
everywhere\cite{Landau75}. Since solutions to (\ref{eqn:gammadiff})
always exist in this case, this proves that the continuum gauge is
indeed a consistent one.

It is important to note that there is actually a significant
additional degree of freedom inherent in the way the decomposition
of $A^\mu$ is made into 4-velocity fields, which goes beyond the
standard gauge freedom. First of all, for each electromagnetic
configuration there will be a continuum of gauge choices for which a
continuum gauge solution set exists. Secondly, for any particular
choice of gauge for which a solution does exist, there will in
general be an entire two-parameter surface of possible solutions for
${\bf v}_+$ and ${\bf v}_-$ at each point in space. We will show
later that these velocity vector fields correspond to the motion of
massive discrete particles, so that this freedom may have a real
physical significance as a possible classical source of dark matter.

\subsection{The Point Charge}

Let us now find the vacuum configuration which describes a positive
charge $q$ positioned at the origin. The corresponding
electromagnetic fields are given by,

\begin{equation}
{\bf E} = {q{\bf{\hat r}}\over r^2} = -\nabla\left({q\over
r}\right)\,,\qquad {\bf B} = 0\,.\label{eqn:eb}
\end{equation}

We seek a 4-potential of the following form which only has
contributions from the motion of the `positive' continuum,

\begin{equation}
A_+^\mu = (\phi_+, {\bf A}_+) = (kc \gamma, k\gamma {\bf
v})\,,\qquad A_-^\mu = (\phi_-, {\bf A}_-) = (-kc, {\bf
0})\,,\label{eqn:seekpot}
\end{equation}

\noindent where the velocity vector field ${\bf v}$ is to be found.
The corresponding electromagnetic fields ${\bf E}$ and ${\bf B}$ are
given by,

\begin{equation}
{\bf E} = -\nabla\phi_+ - {1\over c} {\partial {\bf
A}_+\over\partial t} = - \nabla (kc \gamma) - {1\over c}
{\partial\over\partial t}(k\gamma {\bf v})\,,\qquad{\bf B} =
\nabla\times{\bf A}_+ = \nabla\times(k\gamma {\bf
v})\,.\label{eqn:ebfield}
\end{equation}

For any electrostatic configuration with stationary charges we have
${\bf B} = \nabla\times(k\gamma{\bf v}) = 0$, so there must exist a
scalar field $\psi$ such that $k\gamma{\bf v}=\nabla\psi$. After
some algebraic manipulation this can be seen to imply that,

\begin{equation}
{v\over c} = {\nabla\psi\over\sqrt{k^2c^2 +
(\nabla\psi)^2}}\leq1\,,\qquad\gamma(v) = \left(1 +
{(\nabla\psi)^2\over k^2c^2}\right)^{1/2}\,,\label{eqn:vc}
\end{equation}
so that in terms of $\psi$, the ${\bf E}$ field is given by,

\begin{equation}
{\bf E} = - \nabla\left(\left(k^2c^2 +
(\nabla\psi)^2\right)^{1/2}\right) - {1\over c}
{\partial\over\partial t} \left(\nabla\psi\right)\,.\label{eqn:epsi}
\end{equation}

Because of the rotational and time invariance of the problem, we
need only look for solutions of the form $\psi = \psi(r)$, so that
$\nabla\psi = \partial\psi/\partial r$ and the second term of
(\ref{eqn:epsi}) vanishes. Comparing with (\ref{eqn:eb}), it is
clear that $\psi$ must satisfy,

\begin{equation}
\left(k^2c^2 + \left({\partial\psi\over\partial
r}\right)^2\right)^{1/2} = {q\over r} +
\alpha\,,\label{eqn:psiconst}
\end{equation}

\noindent where $\alpha$ is a constant of integration. Since the
charge is positive and the velocity of the continuum should vanish
at infinity, we require $\alpha = kc$ for a real solution to exist.
From (\ref{eqn:psiconst}), the resulting differential equation for
$\psi$ is as follows,

\begin{equation}
{\partial\psi\over\partial r} = \pm\left(\left({q\over r} +
kc\right)^2 - k^2c^2\right)^{1/2}\,,\label{eqn:dpsidr}
\end{equation}

\noindent where either the positive or negative square root may be
chosen, as the 4-potential depends only on the magnitude of the
velocity and not its direction. There is therefore insufficient
information to specify whether the positive charge acts as a source
or a sink (or both). The solution for the velocity field and the
corresponding Lorentz factor is therefore,

\begin{equation}
{v\over c} = \pm \left(1 - \left(1 + {q\over krc}
\right)^{-2}\right)^{1/2}\,,\quad\quad\gamma=1+{q\over
krc}\,.\label{eqn:vcfinal}
\end{equation}
Note that $q/krc$ becomes singular at the origin, implying that the
continuum velocity in (\ref{eqn:vcfinal}) becomes equal to $c$
there.

The above confirms that the electromagnetic fields outside a
positive point charge can indeed be described by the motion of the
positive continuum, and that the corresponding potential 4-vector
$A_+^\mu$ is expressible in terms of the 4-velocity $u_+^\mu$. An
identical calculation can be performed to show that an analogous
result is true for negative charges.

\subsection{The Plane Electromagnetic Wave}

While in principle one can claim that all electromagnetic
configurations ultimately originate from the presence of charges,
there do exist nontrivial configurations in which no charges are
present, the most obvious and important example being that of the
electromagnetic wave. It is therefore important, both for this
reason and from a historical perspective, to show explicitly how
plane waves arise in the present context from the motion of the
relativistic continuum. We turn to this problem now.

Let us consider a plane electromagnetic wave with wave-vector
$\kappa$ travelling in the $x$-direction with the ${\bf E}$-field
plane-polarised in the $y$-direction. The 4-potential describing
this plane wave is,

\begin{equation}
A^\mu = (0, {\bf A}) = (0, 0, A_y\cos(\omega t-\kappa x), 0
)\,,\label{eqn:planewave}
\end{equation}
(where $\omega=c\kappa$), with corresponding ${\bf E}$ and ${\bf B}$
fields,
\begin{equation}
{\bf E} = (0, E_y, 0) = (0, \kappa A_y\sin(\omega t-\kappa x),
0)\,,\qquad{\bf B} = (0, 0, B_z) = (0, 0, \kappa A_y\sin(\omega
t-\kappa x))\,.\label{planeeb}
\end{equation}
We therefore seek solutions of the form,
\begin{equation}
A_+^\mu = (kc\gamma_+, k\gamma_+{\bf v}_+)\,,\qquad A_-^\mu =
(-kc\gamma_-, -k\gamma_-{\bf v}_-)\,.\label{eqn:pm}
\end{equation}
Applying (\ref{eqn:fourpot}) and equating with (\ref{eqn:planewave})
we obtain the two conditions,
\begin{equation}
\gamma_+=\gamma_-\,,\qquad k(\gamma_+{\bf v}_+-\gamma_-{\bf v}_-) =
(0, A_y\cos(\omega t-\kappa x), 0) \,.\label{eqn:veccond}
\end{equation}

Ignoring equal velocity motions of the `positive' and `negative'
continua which have already been shown to have no electromagnetic
consequences, these conditions allow us to restrict our attention to
solutions of the form,

\begin{equation}
{\bf v}_+ = -{\bf v}_- = (0, v, 0)\,,\qquad\hbox{where}\quad {v\over
c} = {A\over\sqrt{A^2+4c^2}}\,,\label{eqn:soln}
\end{equation}
and we have defined $A=A_y\cos(\omega t-\kappa x)$ for convenience.
The velocities of the positive continuum and the negative continuum
here are equal in magnitude and opposite in direction, so that there
is no net charge, with the motion of both being parallel to the
electric field but $\mp\pi/2$ radians out of phase respectively. It
also follows from (\ref{eqn:soln}) that the velocity of the
continuum can never exceed the speed of light, irrespective of the
intensity of the plane wave. Substituting (\ref{eqn:soln}) into
(\ref{eqn:pm}) the motion of the continuum is given by,

\begin{equation}
u_+^\mu = (\sqrt{c^2/k^2+A^2/4k^2}, 0, A/2k, 0)\,,\qquad u_-^\mu =
(\sqrt{c^2/k^2+A^2/4k^2}, 0, -A/2k, 0)\,.\label{eqn:pmsoln}
\end{equation}

These equations clearly show that the propagation of a plane
electromagnetic wave is described by the oscillation of the medium
in the direction of the electric field - the positive continuum
oscillates $\pi/2$ out of phase with {\bf E} while the negative
continuum oscillates with the same magnitude and precisely the
opposite phase. Thus the propagation of electromagnetic waves is
seen to be a direct manifestation of the oscillations of the
underlying relativistic continuum.

\subsection{Gauge Redundancies and the Principle of Superposition}

While the usual principle of superposition obviously still holds for
the 4-potential, we can now supplement this with the following
continuum-gauge-inspired superposition principle.

Consider two 4-potential fields $A^\mu = k(c\gamma_+-c\gamma_-,
\gamma_+{\bf v}_+-\gamma_-{\bf v}_-)$ and $A'^\mu =
k(c\gamma_+'-c\gamma_-', \gamma_+'{\bf v}_+'-\gamma_-'{\bf v}_-')$ in
the continuum gauge which describe two different 4-velocity field
configurations. Then the superposition of the two field
configurations is described by the 4-potential $A''^\mu =
k(c\gamma_+''-c\gamma_-'', \gamma_+''{\bf v}_+''-\gamma_-''{\bf
v}_-'')$ where the velocity vector field ${\bf v}_+''$ (respectively
${\bf v}_-''$) is given by the pointwise relativistic sum of ${\bf
v}_+$ and ${\bf v}_+'$ (respectively ${\bf v}_-$ and ${\bf v}_-'$),

\begin{equation}
{\bf v}_\pm'' = {{\bf v}_\pm + {\bf v}_\pm'\over1+{\bf
v}_\pm\cdot{\bf v}_\pm'/c^2}\,.\label{eqn:superpose}
\end{equation}

As mentioned earlier, the description of an electromagnetic
configuration in terms of 4-velocities $u_+^\mu$ and $u_-^\mu$ is
far from unique, as for each of the infinite number of 4-potentials
$A^\mu=(\phi,{\bf A})$ with $|\phi|\leq|{\bf A}|$ describing that
particular configuration, there exists an entire two-parameter set
of solutions at each point.

Recall the particular gauge choice in which $\phi=0$ everywhere. We
saw that the simplest `lowest energy' solution is given in this case
by ${\bf v}_+ = - {\bf v}_- = {\bf A}/2k$. However, we also saw that
it is possible to add, relativistically in the sense of
(\ref{eqn:superpose}), the same, arbitrary, possibly time-dependent,
3-velocity vector field to both ${\bf v}_+$ and ${\bf v}_-$ without
changing the 4-potential. If these velocity fields have a real
physical meaning then this additional freedom will correspond to a
large class of vacuum configurations which can perhaps be
interpreted in terms of the motion of an arbitrarily distributed
`Dirac sea' of particles and antiparticles. This provides a means of
adding energy density to the vacuum without any observable
electromagnetic effects.

\subsection{The Continuum as a Massive Relativistic Fluid}

In this section we show that the spacetime continuum must be a
relativistic fluid of massive discrete particles, and that
interactions between sources and sinks give rise to the Lorentz
force equation. The fact that both Maxwell's equations and the
Lorentz force are consequences of the relativistic fluid model is a
strong indication that there is more to this description than mere
formalism, and that classical electrodynamics may in reality have a
fluid dynamical basis.

We saw in (\ref{eqn:vcfinal}) that the velocity of the continuum
decreases with radius outside of the point charge acting as its
source. Had the continuum been massless, its velocity would have
been constant and equal to $c$ everywhere. We therefore conclude
that the continuum has mass and that there is an attractive central
force acting on the continuum outside of the charge.

It is possible to derive an expression for this attractive central
force. In particular, if we assume the charged particle is centred
at the origin, then the force ${\bf f}^i$ acting on an infinitesimal
element of the continuum at radius $r$ must satisfy\cite{Landau75},

\begin{equation}
{\bf f}^i = {d{\bf p}^i\over dt} = m\gamma^3{d{\bf v}^i\over
dt}\,,\label{eqn:relforce}
\end{equation}
where $m=\rho_m \delta V$ is the mass of the test element assuming
that it has mass density $\rho_m$ and occupies volume $\delta V$. To
find the value of $dv/dt$, solve (\ref{eqn:vcfinal}) for $r$ and
differentiate the resulting equation with respect to $t$ to find an
expression for $dv/dt$ in terms of $v$. Rearranging terms and
simplifying, the field at radius $r$ is found to have the form,

\begin{equation}
\gamma^3{dv\over dt} = -{qc\over kr^2}\,.\label{eqn:gravfield}
\end{equation}

Thus there appears to be a Coulombic attraction between the charge
and the continuum around it, with the continuum having a
charge-to-mass ratio of $-c/k$. This is quite mysterious as in our
model charge is defined in terms of the motion of the continuum, so
clearly the continuum itself \emph{cannot} be charged. The mystery
will be resolved in due course.

Assuming continuum conservation, the continuum density $\rho_n$ will
satisfy the following continuity equation,

\begin{equation}
\partial_\mu(\rho_nv^\mu) = {\partial(\rho_n\gamma)\over\partial t}
- \nabla\cdot(\rho_n\gamma{\bf v}) = 0\quad\Rightarrow\quad{1\over
r^2}{\partial\over\partial r}(r^2 \rho_n \gamma v)=0
\,.\label{eqn:radcont}
\end{equation}
where we ignore the time-derivative term as the system is in a
steady state condition, and use the rotational symmetry to rewrite
the divergence term in its spherical polar form. The solution is,

\begin{equation}
\rho_n = {S\over4\pi r^2\gamma v}\,,\label{eqn:contdenst}
\end{equation}
where $S$ is a radius-independent proportionality factor. Now, the
flux of continuum passing through a spherical shell at radius $r$ is
just $\Phi = 4\pi r^2\rho_n\gamma v$ (where the factor of $\gamma$
takes into account to the relativistic contraction in the radial
direction). But this is precisely the constant $S$ in
(\ref{eqn:contdenst}) which can therefore be identified as the
strength of the charged particle sink/source.

\subsection{The Discrete Relativistic Fluid}

We discovered in the previous subsection that there is an
inverse-square law attraction of elements of the continuum towards
the point charge. It is perhaps feasible that the continuum may be
a continuous, compressible, medium whose attractive self-interactions
result in the observed attraction. However, one then has to deal with
divergent contributions to the resultant force on each element of the
continuum from its immediate neighbourhood. These problems can be
avoided by discarding the idea that the continuum is some kind of
continuous elastic medium, but rather consists of a fluid of
interacting discrete particles.

We are therefore led to propose that our relativistic continuum is
a space-filling relativistic fluid and that the electromagnetic
4-potential must be defined in terms of the ensemble motion of the
fluid particles. If the instantaneous fluid velocity at $x^\mu$ is
$\zeta^\mu(x)$, then the 4-velocity appearing in (\ref{eqn:fourvelexp}) is,

\begin{equation}
u^\mu(x) = <\zeta^\mu> = {1\over2}<\zeta_+^\mu+\zeta_-^\mu>\,,\label{eqn:ensemble}
\end{equation}
where $<\zeta^\mu>$ indicates the time-averaged motion of the
particles in the neighbourhood of $x^\mu$ taking into account the contributions from both fluid components. All other electrodynamic
quantities must be defined as time-averages in the same way.

Although the configuration representing a charged particle is in
steady state, the fluid itself remains in constant motion. Recall
that the motion of an individual particle in the co-moving frame of a
relativistic fluid is described by the total derivative\cite{Liboff98},

\begin{equation}
{d\zeta^\mu\over d\tau} = (\zeta^\nu\partial_\nu) \zeta^\mu =
(\zeta_\nu\partial^\nu) \zeta^\mu - {1\over2}
\partial^\mu(\zeta^\nu \zeta_\nu) = - (\partial^\mu \zeta^\nu -
\partial^\nu \zeta^\mu) \zeta_\nu\,,\label{eqn:comoving}
\end{equation}
where we have added a vanishing term using the fact that $\zeta_\mu
\zeta^\mu=c^2$. If we now consider the time-averaged version of
(\ref{eqn:comoving}) and recall the definitions
(\ref{eqn:ensemble}), (\ref{eqn:fourpot}) and
(\ref{eqn:stresstensor}), we find that,

\begin{equation}
k{du^\mu\over d\tau} = - F^{\mu\nu} u_\nu\,,\label{eqn:lorentzfluid}
\end{equation}
which is in the form of the Lorentz force equation. In particular we
find that, on average, each particle moves \emph{as if} it were
charged with $q/m=-c/k$. This is precisely the charge-to-mass ratio
observed in the Coulomb-like attraction of (\ref{eqn:gravfield}),
and so the earlier mystery has been resolved. Because
(\ref{eqn:comoving}) is a basic identity valid for any motion of the
relativistic fluid, this conclusion holds irrespective of the
precise nature of the interactions between the fluid particles.

\subsection{Coulomb's Law and the Lorentz Force Equation}

As further compelling evidence that classical electrodynamics has relativistic
fluid dynamics as its basis, we will now show that the Lorentz force
equation emerges automatically from the interaction between sources
and sinks when they are \emph{not} assumed to be fixed in position.

The integral momentum equation for a fluid tells us that the force
on a target charged particle with charge $Q'$ due to a source
particle of charge $Q$ at distance $r$ is given by the rate of
change of momentum transfer to the target by the particles entering
or leaving the source. If we suppose that the target particle has an
effective radius $R$ then, assuming spherical symmetry, it will have
an effective volume of ${4\over3}\pi R^3$. In accordance with
(\ref{eqn:contdenst}), the density of fluid particles encountering
the target at distance $r$ from the source is $\rho_n(r)$. If we
further assume that each fluid particle is identical with mass $m$,
then the 3-momentum carried by each is given by $m\gamma v$.
Finally, the collision rate will be determined by the strength $S'$
of the target. Thus the force on the target will be given by the
product of these contributions,

\begin{equation}
{\bf F} = {4\over3}\pi R^3 \rho_n m \gamma v S' = {mSS'R^3\over3
r^2} \,,\label{eqn:intmom}
\end{equation}
where he have used (\ref{eqn:contdenst}). This takes precisely the
form of Coulomb's law if we make the following identification,

\begin{equation}
Q = S\sqrt{mR^3\over3}\,,\label{eqn:chargedef}
\end{equation}
where the charge $Q$ is expressed in terms of the strength of the
source $S$, the mass $m$ of the fluid particles and the effective
charge radius $R$. Clearly for (\ref{eqn:intmom}) to hold, positive
charges must effectively act as sources, and negative charges as
sinks, or vice versa. The validity of Coulomb's law in turn
implies the validity of the Lorentz force equation\cite{Einstein20},
as we have assumed from the outset that relativity holds. This
almost completes our description of classical electrodynamics.

The fact that equal velocity contributions of fluid particles from
the positive and negative continua have no electromagnetic effects
means that the net momentum transfer must be zero, which in turn
implies that matter in the negative continuum must have equal and
opposite mass to matter in the positive continuum. All charged particles can then be treated as gravitational sinks of either matter or antimatter, with no need to posit the existence of gravitational sources.

The existence of negative mass particles strongly suggests a
reinterpretation of the 4-potential $A^\mu$ as the net 4-momentum of the positive and negative components of the space-filling fluid. The apparently unphysical negative time signature of the 4-velocity contribution from the negative continuum in (\ref{eqn:fourpot}) can then be attributed to the propagation forwards in time of particles with negative mass, with the constant $k$ being identified as the mass $m$ of the fluid particles. The proposed identification of $A^\mu$ with the fluid's 4-momentum is a radically different interpretation of the 4-potential from the one to which we are accustomed.

\section{The Gravitational Field of a Charged Particle}

In the previous section we have succeeded in deriving the equations
of classical electrodynamics in terms of the motion of a
two-component relativistic fluid where each component is essentially
a time-reversed version of the other. We have not, however,
explained what the origin of these components is, and in that sense
our formulation of classical electrodynamics remains incomplete.

We will show in this section that if, as one would expect, the charged particles have the structure of a classical rotating black hole, then general relativity
actually predicts the existence of the two fluid components, and
this will be sufficient to show that classical electrodynamics is a
necessary consequence of general relativity, given the existence of a space-filling self-gravitating relativistic fluid.

We identify the fluid particles with neutrinos and antineutrinos, and postulate that they are primordial black holes created from the gravitational collapse of gravitational waves. Our model then predicts that the neutrinos will be gravitational dipoles and, following Blanchet\cite{Blanchet06} we show how this can explain the existence of modified Newtonian dynamics as a gravitational polarization effect.

\subsection{The Maximally-Extended Fast Kerr Solution}

The discussion in \S 3.6, and in particular equation
(\ref{eqn:contdenst}) makes clear that the mass density of the fluid
outside a pointlike charged particle tends to infinity as we
approach its centre. This implies that charged particles should be associated with
classical, and in general, rotating, black holes whose spacetime metric is described by the
Kerr solution\footnote{The reader is referred to the monograph by O'Neill\cite{ONeill95} for a detailed study of the Kerr geometry.}.

\begin{figure}
\begin{center}
\includegraphics[scale=0.62]{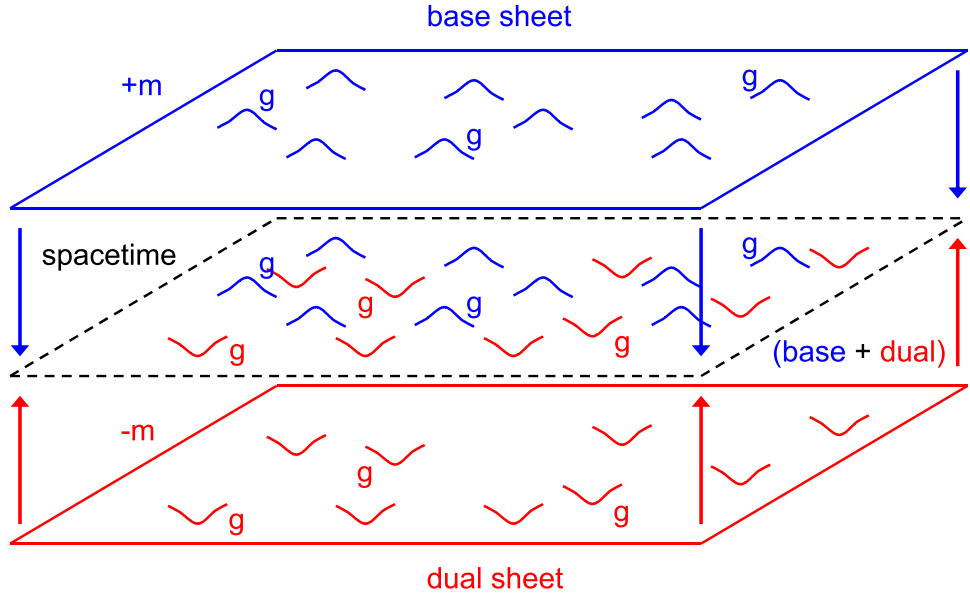}
\caption{The double-sheeted spacetime. Gravitational waves propagate forwards in time on the base sheet (top), and propagate backwards in time on the dual sheet (bottom). Spacetime is a superposition of the dual sheet projected onto the base sheet (middle). The contribution to the curvature from the base sheet is positive, while the contribution from the dual sheet is negative, so that the resulting spacetime is essentially flat.}
\end{center}
\end{figure}

This kind of identification has been proposed by numerous authors in the past because of the striking number of similarities in physical properties between electrons and the charged, rotating, Kerr-Newman solution\cite{Israel70,Arcos04,Burinskii08}. Our model provides an explicit physical explanation for this identification, albeit without the need for a charge as, as we have shown, electric charge is an emergent property due to the motion of the fluid in the neighbourhood of the black hole, and is not an intrinsic physical property of the black hole itself.

We have noted that the two components of the relativistic fluid
constituting the vacuum in our model are time-reversed versions of
each other. This can be explained in the context of the maximally extended fast Kerr solution\footnote{We focus on the fast Kerr solution as for electrons and positrons, with which we will identify the charged particles, as we have $m^2 << a^2$, so that there are no event horizons present.}, which is the uncharged version of the model described in some detail by Arcos and Pereira\cite{Arcos04}. In Kerr-Schild coordinates, the fast Kerr solution takes the form\cite{Hawking75},

\begin{equation}
ds^2=dt^2-dx^2-dy^2-dz^2-{2mr^3\over r^4+a^2z^2}\left({r(x\,dx+y\,dy)-a(x\,dy-y\,dx)\over r^2+a^2}+{z\,dz\over r}+dt\right)^2\,,
\end{equation}
where $r$ is determined up to a sign by $r^4-(x^2+y^2+z^2-a^2)r^2-a^2z^2=0$. Recalling that the maximal-extension of the fast Kerr geometry consists of two four-dimensional spacetime sheets\footnote{The reader is referred to Figure 27 of Hawking and Ellis' monograph\cite{Hawking75} for a clear picture of the double-sheeted maximal extension of the fast Kerr solution, complete with identifications.}, an incoming particle which crosses from the region with $r>0$ (which we will refer to as the `base' sheet) through the wormhole throat at $r=0$, defined by the disc $x^2+y^2<a^2, z=0$, will enter the second region with $r<0$ (which we will refer to as the `dual' sheet). The dual sheet is identical to the base sheet but with the direction of propagation in time reversed. Similarly, any fluid particle in the dual sheet which passes through the wormhole throat at $r=0$ will emerge on the base sheet again with the direction of propagation in time reversed.

The physical picture this implies for a positively (negatively) charged particle is that it acts as a gravitational sink of fluid particles (antiparticles) in the base sheet, and these fluid particles (antiparticles) cross the Kerr wormhole throat and emerge from the dual sheet propagating backwards in time, so that the positively (negatively) charged particle simultaneously acts as a sink for fluid antiparticles (particles) in the dual sheet (see Figure 3). This implies that the 4-potential of a charged particle in \S 3.3 will have an equal contribution from both spacetime sheets on either side of the Kerr wormhole.

As explained by Chardin\cite{Chardin97}, particles travelling `backwards' in time in the dual sheet will continue to interact with particles in the base sheet, and in particular, will appear to have the same physical properties as antiparticles travelling forwards in time on the base sheet\footnote{Note that Chardin also associates a negative gravitational mass with antimatter, and provides additional independent evidence for this conclusion.}. The base sheet and its dual will therefore appear to be superimposed onto a single four-dimensional spacetime, and to an external
observer the process of absorption and reflection of a fluid particle will look not like a
single infalling particle, but like a particle-antiparticle
annihilation event occurring at the wormhole throat. This description is
consistent with the analysis of Hadley\cite{Hadley02} who concludes
that the failure of time-orientability of a spacetime region would
be indistinguishable from a particle-antiparticle annihilation
event.

\begin{figure}
\begin{center}
\includegraphics[scale=0.62]{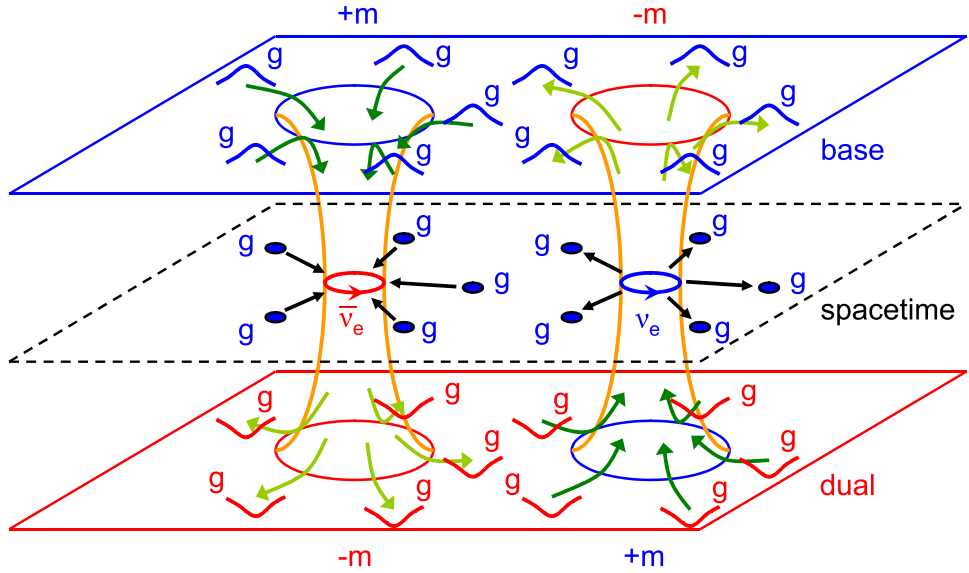}
\caption{Neutrinos as sinks of gravitational waves. Neutrinos (antineutrinos) are formed from the gravitational collapse of gravitational waves in the dual (base) sheet, and have the spinorial geometry of a maximal fast Kerr wormhole throat connecting the base sheet to the dual sheet. The resulting dipole structure means that the isolated neutrinos and antineutrinos are massless but become polarised in a gravitational field. To an external observer, antineutrinos act as sinks of gravitational waves, while neutrinos act as sources.}
\end{center}
\end{figure}

An interaction between two charged particles can then be described by the exchange of a photon, which can be pictured as a twisted closed timelike loop formed by a fluid particle leaving the source particle\footnote{The fluid particles and antiparticles will be identified in the next section as gravitational dipoles, and are hence polarisable. They will therefore attract each other through a Van der Waals type interaction during their passage from the source charge to the target, resulting in general in a helical trajectory, so that the closed photon loop appears to be twisted.}, reflecting off the target particle and returning to the source, where it is once again reflected to return to its original position and time orientation (see Figure 4). This does not imply any causal
inconsistency, as the process would merely have the appearance of a
pair creation event at the source charge followed by a subsequent pair annihilation event at the target charge. As indicated in \S 3.8, whether the interaction results in an attraction or repulsion will depend on the nature (i.e. matter on antimatter) of the source and target charges, as well as the type of particles exchanged.

There exists a natural time-reversal symmetry in this model associated with the swapping of the base and dual spacetime sheets, which also exchanges the identity of matter and antimatter, and hence the sign of the gravitational mass (see Figure 1). This overall time-symmetric picture of charges and their interactions is reminiscent of the Wheeler-Feynman absorber theory\cite{Wheeler45} of radiation, as well as Cramer's transactional interpretation\cite{Cramer86} of quantum mechanics, and also of Hoyle and Narlikar's action-at-a-distance\cite{Hoyle95} cosmology. These connections will be explored in more detail in future work.

\subsection{Neutrinos as Gravitational Dipoles}

Hadley has shown\cite{Hadley00} that geon-like elementary
particles in classical general relativity, of which our charged
particles are particular examples, will naturally have the
transformation properties of spinors if the spacetime manifold is
not time-orientable. We conclude that the charged particles in our model are spinors, and we now formally identify them with electrons and positrons\footnote{Conversely, it is natural to
propose that all of the elementary fermions observed in nature are
gravitational solitons corresponding to stable topological
configurations of classical black holes.} (see Figure 3).

\begin{figure}
\begin{center}
\includegraphics[scale=0.62]{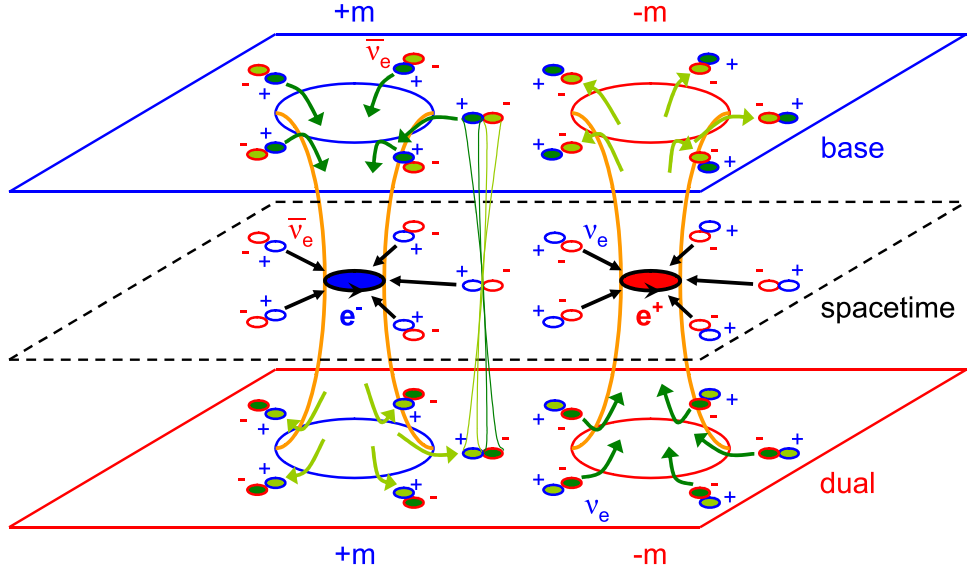}
\caption{Electrons as sinks of antineutrinos. Electrons (positrons) are formed from the gravitational collapse of antineutrinos (neutrinos), and like neutrinos they have the geometry of the maximal fast Kerr solution. The antineutrinos (neutrinos) become polarised in the gravitational field of the charges, giving rise to the electron's (positron's) positive (negative) mass. In addition, the motion of the antineutrinos (neutrinos) in the neighbourhood of the Kerr wormhole throat give rise to the electron's (positron's) negative (positive) charge.}
\end{center}
\end{figure}

Given their fundamental nature, the fluid particles can be assumed to be ground
state gravitational solitons formed from the collapse of intense
gravitational waves, which are themselves nothing but ripples in
spacetime.
It is fairly well-established
that gravitational waves of sufficient intensity can collapse to
form black holes\cite{Abrahams92,Alcubierre00}, so presumably there
was enough energy in the early universe for these primordial black
holes to be formed in enormous quantities\footnote{It is generally assumed that primordial black holes must be at least of the order of the Planck mass, and that low mass black holes would rapidly evaporate through emission of Hawking radiation. However both of these conclusions are based upon more fundamental assumptions regarding the validity and applicability of quantum theory at these scales, and will not necessarily hold if these assumptions are incorrect\cite{Helfer03}. In particular, such assumptions do not hold in the present model, which is purely classical in nature and in which Kerr solutions are stable as the set of geodesics encountering the singularity is of measure zero. It will be shown in a future work that quantum theory emerges as a consequence of the the non-causal structure of the Kerr geometry.}. Having the geometry of the maximal fast Kerr solution,
the fluid particles will themselves transform as spinors. Our earlier picture of fluid particles, which had assumed that each particle inhabits either the base sheet or its dual, will therefore need to be extended to allow each particle to have one component (i.e. the two regions on either side of the wormhole throat) on each spacetime sheet.

We know that the fluid particles are responsible, through their
motion, for the appearance of charge, and so cannot be charged
themselves. The only uncharged spinors observed in nature are the
neutrinos, which do indeed fill spacetime, so it is natural to
identify our fluid particles and antiparticles with neutrinos and
antineutrinos respectively. The identification of these microscopic classical black holes with neutrinos seems
particularly appropriate as Einstein and Rosen suggested the
same identification themselves in their original
paper\cite{Einstein35}. The two halves of the Einstein-Rosen bridge,
one on each spacetime sheet, will have equal and opposite mass\cite{Arcos04}, and
the neutrino therefore has the structure of a gravitational dipole (see Figure 2).

%
%

Because they are space-filling, the neutrinos effectively act like `pegs'
holding together the two spacetime sheets of the double-sheeted `Dirac-Milne' universe recently described by Benoit-L\'{e}vy and Chardin\cite{Benoit08}. We saw
in \S 3.4 that the propagation of electromagnetic waves could be
represented by the oscillations of the positive and negative
momentum particles out of phase. This implies that
electromagnetic waves are described by the bounded oscillatory
motion of neutrinos and antineutrinos, which is strongly reminiscent of the
`neutrino theory of light' that originated with the ideas of de
Broglie\cite{DeBroglie32}. The neutrinos, then, are responsible for
the physical vacuuum acting as a `luminiferous aether'.

\begin{figure}
\begin{center}
\includegraphics[scale=0.6]{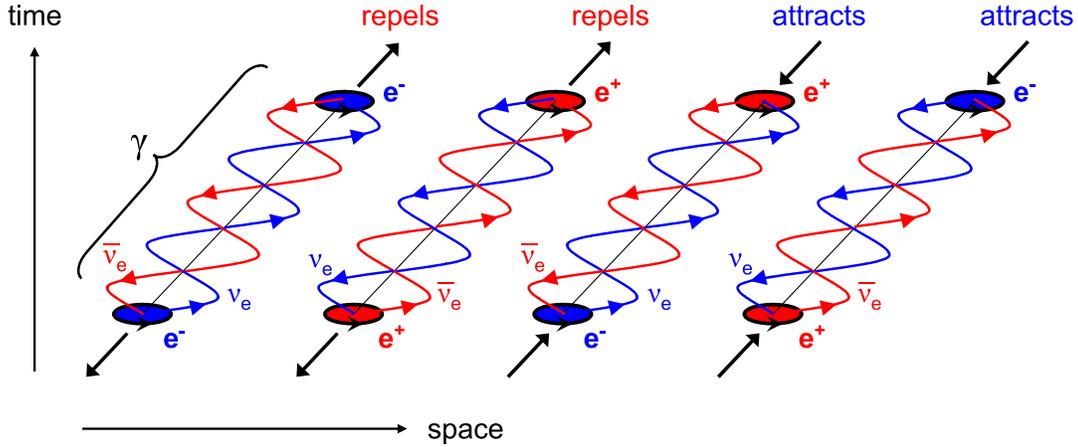}
\caption{Like charges repel while opposite charges attract. Source electrons (positrons) emit neutrinos (antineutrinos) which are reflected back to the source by the target charge. The momentum transferred by the neutrinos cause target electrons (positrons) to be repelled (attracted), while antineutrinos cause target electrons (positrons) to be attracted (repelled). The photons look like twisted closed timelike loops formed by neutrino-antineutrino pairs travelling from source to target. The neutrino and antineutrino polarise each other, resulting in a van der Waals type interaction which is responsible for their oscillatory motion during the traversal from source to target.}
\end{center}
\end{figure}

The cancellation of the contributions from the base and dual components constituting the neutrino means that an isolated
neutrino would be expected to have zero mass. However the
gravitational dipole structure of the neutrino means that it will
become polarised in the presence of a gravitational field in analogy
with electrodynamics, and this polarisation will be manifested by a
small relative translation of the two spacetime sheets with a magnitude of the same order as the radius of the ring singularity associated with the neutrino. As a result,
the neutrino will be attracted towards any nearby massive body,
irrespective of whether that body consists of matter or antimatter.
This will give the neutrino the appearance of a small but finite
mass of the same sign as the source.

This property can explain why electrons, which act as sinks of antineutrinos, appear to have mass, even though the antineutrinos themselves are massless. As the antineutrinos are attracted closer to the Kerr singularity, they become increasingly polarised, giving rise to the electron's positive mass (see Figure 3). Similarly, positrons will attract neutrinos with opposite polarisation, giving rise to the positron's negative mass.

The matter content of our model, which makes use of some of the results of \S 5, is summarised pictorially in Figures 1-4.

\subsection{Exotic Cold Dark Matter and Modified Newtonian Dynamics}

The polarisability of the neutrino means that in a matter-dominated region of the universe neutrinos will appear to
have a small positive mass which would be compatible with recent
observations. Despite their small apparent mass, it is natural to
conjecture that the sheer number of neutrinos which fill spacetime
could potentially account for the apparent missing dark matter in
the universe.

The other significant physical consequence of the polarisability of
the neutrino is that the physical vacuum will act like a
gravitational dielectric or `digravitic' in analogy with dielectrics
in electrodynamics and this turns out to be the key to understanding
the presence of modified Newtonian dynamics.

Indeed Blanchet showed in a recent paper\cite{Blanchet06} that if
there were to exist a space-filling `aether' consisting of exotic
dark matter particles which take the form of matter-antimatter
dipoles (antimatter having negative mass), then this would
satisfactorily explain the existence of MOND as a simple
gravitational polarisation effect.

Clearly our model of the physical vacuum fits this description
perfectly, with the neutrinos playing the role of the exotic matter
which Blanchet describes. The implication of Blanchet's results is
that our model, which itself is based purely on Einstein's general
theory of relativity, actually predicts the existence of modified
Newtonian dynamics.

To show how this works, following Blanchet, let us denote the
gravitational potential by $U$. Then the gravitation field ${\bf g}$
associated with it is,

\begin{equation}
{\bf g} = \nabla U\,.\label{eqn:gravityfield}
\end{equation}

Let the spatial vector ${\bf d}$ denote the separation between the
positive and negative mass components of the (anti)neutrino, which will vary
with the strength of the gravitational field. Then the dipole moment
associated with each (anti)neutrino is,

\begin{equation}
{\bf p} = m {\bf d}\,.\label{eqn:dipole}
\end{equation}
If the number density of dipoles is $n$, then the gravitational
polarisation ${\bf P}$ will be,

\begin{equation}
{\bf P}=n{\bf p}\,.\label{eqn:gravpol}
\end{equation}

The positive gravitational mass component of the neutrino (we assume that
all particles have positive inertial mass) will always be attracted
by an external mass distribution consisting of ordinary matter,
while the negative mass component will be repelled. The orientation of the
dipole will then be such that the dipole moment ${\bf p}$, and hence
the polarisation ${\bf P}$ points in the direction of the
gravitational field ${\bf g}$.

The MOND equation in the form derivable from a non-relativistic
Lagrangian is,

\begin{equation}
\nabla\cdot(\mu{\bf g})=-4\pi G\rho\,,\label{eqn:mond}
\end{equation}
where $\rho$ is the density of ordinary matter, and the Milgrom
function $\mu$ depends on the ratio $g/a_0$ where $g=|{\bf g}|$ is
the magnitude of the gravitational field and $a_0$ is the constant
acceleration scale. The MOND regime corresponds to the limit of weak
gravity when $g\ll a_0$, in which case $\mu(g/a_0)\approx g/a_0$.
Similarly, in the strong field Newtonian regime when $g\gg a_0$,
$\mu(g/a_0)\rightarrow1$, and we recover Newton's law.

To make the analogy with electrostatics clear, note that the
equation for an electric field in a dielectric medium
is\cite{JDJ98},

\begin{equation}
\nabla\cdot[(1+\chi_e){\bf
E}]=\rho_e/\epsilon_0\,,\label{eqn:dielectric}
\end{equation}
where $\chi_e$ denotes the electric susceptibility of the medium and
depends on the magnitude of the electric field. Typically
$\chi_e>0$, which corresponds to screening of the electric charges
by the dielectric. The electric polarisation is then defined by,

\begin{equation}
{\bf P}_e=\chi_e\epsilon_0{\bf E}\,,\label{eqn:epol}
\end{equation}
In the case of gravitation, we can write the Milgrom function
$\mu(g/a_0)$ as,

\begin{equation}
\mu=1+\chi\,,\label{eqn:gravsus}
\end{equation}
where $\chi=\chi(g/a_0)$ is the gravitational susceptibility of the
digravitic medium. The corresponding gravitational polarisation
${\bf P}$ is then,

\begin{equation}
{\bf P}=-{\chi\over4\pi G}\,{\bf g}\,,\label{eqn:gp}
\end{equation}

Since in the gravitational case ${\bf P}$ is in the same direction
as ${\bf g}$, the gravitational susceptibility $\chi$ must be
negative,

\begin{equation}
\chi<0\,,\label{eqn:negsus}
\end{equation}
which is compatible with the MOND prediction that $0<\mu<1$ which
requires that $-1<\chi<0$. The underlying reason for the negative
gravitational susceptibility is simply the fact that like masses
attract whereas like charges repel.

The equations of motion for the positive and negative mass
components of the dipole are as follows,

\begin{equation}
m{d^2{\bf x}_1\over dt^2}=m{\bf g}({\bf x}_1)-{\bf f}({\bf x}_1-{\bf
x}_2)\,,\label{eqn:planckone}
\end{equation}
\begin{equation}
m{d^2{\bf x}_2\over dt^2}=-m{\bf g}({\bf x}_2)+{\bf f}({\bf
x}_1-{\bf x}_2)\,,\label{eqn:plancktwo}
\end{equation}
where ${\bf x}_1$ and ${\bf x}_2$ are the centre-of-mass positions of the positive and negative mass components respectively, and ${\bf f}({\bf x})$ is the force
between them as a function of their separation. Let us transfer to
new coordinates ${\bf x}={1\over2}({\bf x}_1+{\bf x}_2)$
representing the centre of the dipole and the dipole moment ${\bf
p}=({\bf x}_1-{\bf x}_2)$. Then after a first order Taylor expansion
of ${\bf g}({\bf x})$, the evolution equation for the dipole is
found to be,

\begin{equation}
{d^2{\bf p}\over dt^2} = 2m{\bf g}-2{\bf f}+{\mathcal
O}(d^2)\,,\label{eqn:dipevo}
\end{equation}
while the equation of motion for the dipole in a gravitational field
is,

\begin{equation}
2m{d^2{\bf x}\over dt^2} = ({\bf p}\cdot\nabla)\nabla U + {\mathcal
O}(d^2)\,.\label{eqn:dipmotion}
\end{equation}

This tells us that the motion of the dipole is governed not by the
strength of the gravitational field, but by its gradient, namely the
tidal gravitational field. This means that the dipole will remain
stationary in a constant gravitational field, and in a gravitational
field outside a spherical massive body with potential $U\sim 1/r$,
the dipole's acceleration will be of the order of $1/r^3$ instead of
the usual $1/r^2$ for an ordinary particle. Clearly the neutrinos
seem to violate the equivalence principle, having an inertial mass
of $2m$ and a gravitational mass of zero, and as such are good
candidates for cold dark matter.

The question remains as to how the the dipole separation ${\bf d}$
varies with the field strength ${\bf g}$. Unlike Blanchet, we have
no need to postulate a new internal force of non-gravitational
origin, as we are well aware that what physically is happening when
a neutrino is polarised is an attempt to rotate and pull the two poles into alignment with the external field. Small perturbations
can be expected to follow a linear Hooke's law pattern, as evidenced
by the quasiharmonic motion describing the propagation of
electromagnetic waves derived in (\ref{eqn:soln}), which represents
a similar physical process. However there would be expected to be an
asymptotic value beyond which the two poles can no longer be
rotated or stretched, and so a reasonable parametric form for the dipole
separation may be as follows,

\begin{equation}
d(g) = d_0\tanh(\alpha g)\,,\label{eqn:dipsep}
\end{equation}
where $d_0$ is the dipole separation at saturation in the strong
field limit, and the Hooke's law `spring constant' is given by
$\alpha d_0$. From (\ref{eqn:gp}) and (\ref{eqn:dipsep}) the
gravitational susceptibility $\chi$ would then take the form,

\begin{equation}
\chi=-4\pi Gnm{d\over g}\,,\label{eqn:susg}
\end{equation}
so that the Milgrom function becomes approximately,
\begin{equation}
\mu(g/a_0)\approx 1-4\pi Gnm{d_0\tanh(\alpha g)\over
g}\,.\label{eqn:milg}
\end{equation}
This has the correct property $\mu\rightarrow 1$ in the Newtonian
regime when $g\gg a_0$. In the MOND regime, corresponding to the
limit $g\rightarrow0$, we require $\mu=g/a_0+{\mathcal O}(g^2)$ in
order to explain the flat rotation curves of galaxies. This fixes
the value of $\alpha$ in (\ref{eqn:dipsep}),

\begin{equation}
\alpha = {1-{g/a_0}\over4\pi Gnmd_0}\,.\label{eqn:tanhconst}
\end{equation}

It seems unlikely that the true dependence of the dipole separation
on the gravitational field strength will vary significantly from
(\ref{eqn:dipsep}), and that any differences are likely to have
limited physical consequences. The essential features that need to
be present are that $d'(g)/g$ in the zero field limit agrees with
observations, and that $d/g\rightarrow0$ in the strong field limit.

The physical picture we then have is as follows. When there is no
gravitational field there is no polarisation, while at small but
finite gravitational fields, the polarisation of the vacuum
increases linearly with field strength, corresponding to the MOND
regime. As the field strength increases further, the polarisation
becomes saturated, reaching an asymptotic value, so that eventually
the effects of vacuum polarisation become negligible in comparison
with the external field and we return to the usual Newtonian regime.
All of this appears to be a
nontrivial consequence of classical general relativity without
modification and without needing to introduce any new particles not
already observed in nature. Indeed it appears that neutrinos and
antineutrinos themselves can be identified as the `missing' cold
dark matter.

\section{Antigravity and its Cosmological Consequences - Some Speculations}

The possibility of negative mass in the context of general
relativity was first discussed by Bondi\cite{Bondi57}. The article
by Nieto and Goldman\cite{Nieto91} reviews theoretical arguments
against the existence of antigravity. Nevertheless there has been a
renewed interest in the possibility of
antigravity\cite{Hossenfelder05,Ni03} on account of recent
cosmological observations, including proposals for experimental
verification\cite{Chardin97,Hajdukovic06}. Chardin has argued that the existence of antigravity could explain the observed CP violation in the neutral kaon system\cite{Chardin97}. Moreover, antiparticles are
predicted to have negative mass by the Dirac equation in
relativistic quantum theory, and negative mass regions are actually rather commonplace in solutions of the Einstein equations in general relativity. These latter two facts seem in themselves to be sufficient
reason to take the idea of antigravity quite seriously, rather than considering it to be an inconvenience or an embarrassment that must be explained away or brushed underneath the carpet and simply ignored.

We have shown here that antigravity must exist in the classical
realm without invoking quantum mechanics, and that classical
electrodynamics emerges directly from general relativity as a
result. The presence of antigravity would naturally be expected to have consequences for
some of the major outstanding issues in cosmology, and we very
briefly discuss a number of such speculations here, most of which have already
been put forward by others in different contexts.

\subsection{Matter Dominated Regions and the Accelerating Expansion
of the Universe}

Perhaps the simplest consequence of antigravity is that matter will
tend to clump together with matter and antimatter will tend to clump
together with antimatter, but the two types of matter will repel. If
we assume that there are comparable amounts of both matter and
antimatter in the universe, the result will be region(s) of space
which are matter dominated, and other region(s) which are antimatter
dominated. In particular we appear to live in a matter-dominated
region of the universe.

The repulsion between matter and antimatter dominated regions should
in principle be observable, and indeed in the case of an open
universe it would predict that the universe should expand at an
accelerating rate, as is observed. This was mentioned by both
Ripalda\cite{Ripalda99} and Ni\cite{Ni03}. Ni goes on to suggest
that the supernovae observed to undergo acceleration may do so
because they consist of antimatter and there is a repulsive force
exerted upon them by inner galaxies consisting mainly of matter. He
also proposes that increasing antimatter domination is responsible
for the increasing rate of star formation at increasingly remote
distances.

On the other hand it is a curious coincidence that the observed size
of the universe is very close to the size that would be expected for
a black hole of the same mass. If the observable universe is indeed
enclosed within a non-traversable event horizon, or is otherwise
bounded, then the result of antigravity would be a universe which
undergoes cycles of expansion and contraction\cite{Hossenfelder06}.
If that is the case, then clearly we are in an accelerated expansion
phase following an earlier deceleration, and this would agree with
cosmological observations\cite{Padmanabhan06}.

\subsection{Gamma Ray Bursts}

If matter and antimatter dominated regions do exist as antigravity
would predict, then wherever the boundaries between the matter and
antimatter dominated regions meet there will be some `rubbing
together' of the two, resulting in massive particle-antiparticle
annihilation events which will give off huge bursts of
electromagnetic radiation. This `cosmic lightning' would be observed
as gamma ray bursts. This explanation for the origin of gamma ray
bursts has also been suggested by Ripalda\cite{Ripalda99}.
Ni\cite{Ni03} further argues that the Earth is actually near the
centre of a matter-dominated region based upon the observed
isotropic distribution of gamma ray bursts.

\subsection{The Cosmological Constant and Spacetime Curvature}

Einstein's general theory of relativity allows for the presence of a
cosmological constant, but the value observed for $\Lambda$ is over
120 orders of magnitude smaller than that predicted from quantum
vacuum effects. The tiny value observed for the cosmological
constant can be explained at a fundamental level in the context of our model
if all particles on the dual sheet have opposite energy to those on the
base sheet. Because these particles always occur in pairs, their
contribution to the vacuum energy will cancel, resulting in no net
contribution to the cosmological constant. The small value of the
cosmological constant could then be attributed to asymmetries
between the base sheet and its dual, which in the context of our
model can only be attributed to the presence of excess gravitational
waves on the base sheet, namely those which do not collapse to form
elementary particles such as neutrinos. Quiros\cite{Quiros04} has also suggested that there may
exist two vacua, one gravitating and one antigravitating resulting
in the mutual cancellation of their contributions to the
cosmological constant. Alternatively, Moffat\cite{Moffat96} and
Padmanabhan\cite{Padmanabhan04} propose that fluctuations of vacuum
energy density may be responsible for the observed cosmological
constant.

The universe appears to be approximately flat with only a small
positive curvature on cosmic scales. If energy is associated with
curvature, then the same considerations above would explain the
relative flatness of the universe, with the uncollapsed
gravitational waves accounting for any small positive curvature that
remains.

\subsection{The Dirac-Milne Cosmology}

The picture that emerges from our model is of a time-symmetric double-sheeted universe which treats matter and antimatter with an equal status. The compatibility of our model with the Wheeler-Feynman absorber theory of radiation  suggests a reappraisal of the quasi-steady-state cosmology proposed by Hoyle and Narlikar\cite{Hoyle95}.

More recently, Benoit-L\'{e}vy and Chardin, who also suggest the identification of elementary particles with the fast Kerr geometry, have examined the properties of a matter-antimatter-symmetric Milne spacetime filled with Kerr-Newman type particles, which they refer to as the `Dirac-Milne' cosmology, and have found an excellent agreement with cosmological data without the high level of manual fine-tuning of parameters required with the $\Lambda$-CDM standard model\cite{Benoit09}. In particular, the Dirac-Milne cosmology appears to satisfy the cosmological tests for the age of the universe, big bang nucleosynthesis, type Ia supernovae data and even provides the degree scale for the first acoustic peak of the cosmological microwave background.
%

\subsection{The Relationship Between Energy, Mass and Curvature}

Our model predicts that there should be two superimposed spacetime
sheets - the `base' sheet, and the dual `sheet'. In the dual sheet,
time goes in the opposite direction relative to the base sheet, so
that an observer on the base sheet will observe a particle on the
dual sheet to be travelling backwards in time. There must also be
both particles and antiparticles on the same sheet, with the
antiparticles travelling in the opposite direction in time to the
particles.

However all of these identifications are relative to the particular
frame of reference used, and different observers will in general
disagree about what constitutes matter or antimatter and which sheet
is the base sheet and which is its dual. Let us select one
particular observer arbitrarily in order to establish a convention.
According to that observer, there are four types of matter in
existence, namely, matter on the base sheet ($B^+$), antimatter on
the base sheet ($B^-$), matter on the dual sheet ($D^+$), and
antimatter on the dual sheet ($D^-$).

Now, in addition to (a) the direction of propagation in time, there
is associated with all of these types of matter, (b) a gravitational
mass, (c) an inertial mass, (d) an energy, and (e) an apparent
curvature of the surrounding spacetime. We would like to find out
the sign of each of these five parameters for each of the four
matter types in the context of our model. Based upon known
observations we can come to the following conclusions:

\begin{itemize}

\item To account for Coulomb's law in electrodynamics, it  must be
the case that particles and antiparticles on the same sheet have
opposite gravitational mass. This has already been discussed in \S
3.8.

\item To account for the zero mass of isolated neutrinos, which have
a dipole structure, as well as for the existence of modified
Newtonian dynamics, which is a consequence of gravitational
polarisation of neutrinos, matter in the base sheet must have
opposite gravitational mass to antimatter in the dual sheet.

\item Because we live in a matter dominated part of the universe, and
also indirectly from the observation that the universe is expanding
at an accelerating rate, matter and antimatter in the same sheet
must repel. This, together with their known sign of gravitational
mass implies that the inertial mass of all particle types is
positive.

\item Because matter attracts matter and antimatter attracts
antimatter, both of these must be associated with positive curvature
in the same spacetime sheet as the observer.

\item The near flatness of spacetime means that the total curvature
is close to zero. This means that the curvature associated with
matter and antimatter on the dual sheet must be negative and cancel
the curvature due to matter and antimatter on the base sheet.

\item Energy is released when matter and antimatter annihilates, so
conservation of energy requires that matter and antimatter on the
same sheet have the same sign of energy.

\item The tiny value of the cosmological constant
implies that the total energy of the vacuum must be very small, so
that the energy of both matter and antimatter in the dual sheet must
be negative.

\end{itemize}

\noindent Putting all of this information together, we finally arrive at the
following table:

\vspace{5mm} \hspace{3cm}
\begin{tabular}{|c|cccc|}
\hline Particle type & $B^+$ & $B^-$ & $D^+$ & $D^-$\\
\hline Direction of time & $+$ & $-$ & $-$ & $+$ \\
Gravitational mass & $+$ & $-$ & $+$ & $-$ \\
Inertial mass & $+$ & $+$ & $+$ & $+$ \\
Energy & $+$ & $+$ & $-$ & $-$ \\
Spacetime curvature & $+$ & $+$ & $-$ & $-$ \\
\hline
\end{tabular}
\vspace {5mm}

We see from the table that energy is associated with spacetime curvature, and
that neither of these are equivalent to either gravitational or
inertial mass. Furthermore, we see that the principle of equivalence
does not strictly hold for antimatter, as these have opposite (as
opposed to equal) inertial and gravitational masses, requiring that
the principle be generalised to describe antimatter correctly. The
same conclusion was reached by Hossenfelder\cite{Hossenfelder06}.

\section{Discussion and Summary}

We began with a simple description of the physical vacuum as a relativistic
fluid in motion. We then showed that it was possible to derive classical
electrodynamics in terms of the motion of a two-component matter-antimatter
fluid with the electromagnetic 4-potential being identified with the time-averaged
4-momentum of the fluid. Charged particles act as gravitational sinks, and were therefore
assumed to have a maximal fast Kerr geometry, which implies that spacetime is
double-sheeted, and that antimatter has negative mass. The fluid particles were
postulated to be neutral spinors formed from gravitational collapse of gravitational
waves, and were identified as neutrinos. These have the same maximal fast Kerr
geometry, which in turn implies that they are gravitational dipoles. Because the neutrinos are
space-filling, this means that the vacuum must be gravitationally polarisable, and
this was shown to be sufficient to explain the existence of modified Newtonian
dynamics.

The entire model can be derived essentially from first principles from the
general relativistic treatment of a space-filling fluid, and hence we have been
able to show that classical electrodynamics and modified Newtonian dynamics are both
non-trivial consequences of general relativity. The time and matter-antimatter symmetry
of the model is compatible with the Wheeler-Feynman absorber theory of radiation, and
hence Hoyle and Narlikar's action-at-a-distance cosmology. The non-causal structure of
the maximal fast Kerr solution gives rise to time-reversing process which suggest a
relationship with Cramer's transactional interpretation of quantum mechanics. It will be
interesting to see whether this connection with quantum theory can be made explicit so that
the model is able to provide a sound physical basis for quantum theory.

In terms of possible cosmological consequences we have discussed briefly how the model's prediction of antigravity may help us to understand the
accelerating expansion of the universe, the origin of gamma ray bursts, the
smallness of the cosmological constant, the relationship between
mass, energy and curvature, and also how the model may give rise to a Dirac-Milne universe
which seems to agree well with cosmological data without the need for fine-tuning of parameters.
Trayling and Baylis\cite{Trayling01}
were able to derive the standard model gauge group in terms of Clifford algebra on a $7+1$-dimensional spacetime, so it would be of interest to see whether the standard model gauge group can similarly be derived from our model as a consequence of spacetime being double-sheeted.

\section{Acknowledgements}

I would like to thank Jan Bielawski, Steve Carlip, Steve Gull, Hugh Jones, Daryl McCullough,
Abhas Mitra and Tom Roberts for useful discussions and feedback, and Roy Pike for
his valuable support and encouragement. I would like to dedicate
this paper to John Peach and Simon Altmann, my ex-tutors at
Brasenose College.

\end{document}